\documentclass[epj]{svjour}
\usepackage{graphics}
\usepackage{graphicx}
\usepackage{epsfig}
\usepackage{ulem}
\usepackage{morefloats}
\usepackage{rotating}
\usepackage{color,xcolor}
\usepackage{soul}
\usepackage{amsmath}
\usepackage{amssymb}
\usepackage{url}
\usepackage[utf8]{inputenc}

\begin{document}
\title{Strong magnetic fields: neutron stars with an extended inner crust}
\author{Helena Pais\inst{1} \and Bruno Bertolino\inst{1} \and Jianjun Fang\inst{2} \and Xiaopeng Wang\inst{2} \and Constan\c ca Provid{\^e}ncia\inst{1}}% etc
%}                     % Do not remove
\mail{Helena Pais, \texttt{hpais@uc.pt}, Constan\c ca Provid\^encia, \texttt{cp@uc.pt}}
\institute{CFisUC, Department of Physics, University of Coimbra,
   3004-516 Coimbra, Portugal. \and School of Physics and Physical Engineering, Qufu Normal University, 273165 Qufu, China.}
\date{Received: date / Revised version: date}
\abstract{
Using relativistic mean-field models, the
formation of clusterized matter, as the one expected to exist in the
inner crust of neutron stars, is determined under the effect of strong magnetic
fields.  As already predicted from a calculation of the unstable
modes resulting from  density  fluctuations at subsaturation densities, we confirm in the present work
that for magnetic field intensities of the order of $\approx
  5\times10^{16}$ G to $5\times 10^{17}$ G, pasta phases may occur for densities well above the zero-field crust-core transition density. This confirms that the extension of the
crust may be larger than expected. It is also verified that the
equilibrium structure of the clusterized matter is very sensitive to
the intensity of the magnetic fields. As a result, the decay of the
magnetic field may give rise to internal stresses which may result on
the yield and fracture of the inner crust lattice.
%
%\PACS{
%      {PACS-key}{discribing text of that key}   \and
%      {PACS-key}{discribing text of that key}
%     } % end of PACS codes
} %end of abstract
\maketitle
\section{Introduction}
\label{intro}

Neutron stars are compact objects with a very complex structure. Their
interior is generally divided into an outer and  inner crust and an outer and
inner core. The information coming to the surface  from the interior of the star will
necessarily cross the crust, and, therefore, it is essential to know
well the constitution of the crust not only because of the processes
that  result directly from the crust but also in order to learn about
the core of the star.
The inner crust that lies directly at the border with the core  has a
complex constitution. In particular, it is generally accepted that  the layers just below the
crust-core transition are formed by frustrated matter  resulting from the
competition between  the 
strong and the electromagnetic interactions \cite{Ravenhall-83,Hashimoto-84,Horowitz-05,watanabe2005,Maruyama2005,Avancini-08,Avancini-10,pais2012PRL,Bao-14}.  Neutron star phenomena
that seem to be directly related with the properties of the inner
crust are glitches \cite{Link-99,Chamel-13,Andersson-12} or the evolution of
the magnetic field intensity  \cite{Pons-13}.
Neutron superfluidity in the inner crust plays an essential role
in the interpretation of the glitches \cite{Link-99}, but the
entrainment of neutrons to the solid crust may raise some problems on
whether the description of glitches is totally described by the crust
\cite{Chamel-13,Andersson-12}. In \cite{Pons-13}, the authors have shown that
the  dissipation of the magnetic field responsible for a fast spin
down of the stars could be a signal of the presence of a highly
resistive layer of matter in the inner crust, and that the existence of pasta phases as
suggested in \cite{Ravenhall-83}, were a possibility.

A class of  neutron stars known as magnetars is characterized by very
intense surface magnetic  fields which span the range $\sim
10^{12-15}$ G \cite{kaspi14,pulsars}. Inside these stars the magnetic field could be even more
intense, and several estimates and simulations suggest that fields  of
the order of  $\sim 10^{18}$ G  could exist in their interior \cite{lai1991cold,cardall2001effects,Broderick02,Chatterjee15,Gomes19,Sengo20}.

In the present study, we want to calculate the structure of the bottom
layers of the inner crust in the presence of a strong magnetic
field. It has been reported that the extension of the inner crust under
these conditions could be much larger, and it could have a complex structure
constituted by alternating regions of clusterized and non-clusterized
matter above the $B=0$ crust-core transition \cite{Fang16,Fang17,Chen17,Fang17a},
defining a transition region with a finite width.

The extension of the crust in a neutron star is directly connected to
the presence of a liquid-gas type transition in nuclear matter
\cite{Mueller-95}.  In Refs.~\cite{Fang16,Fang17}, the effect of a strong
magnetic field on 
the density that characterizes the transition to
homogeneous matter  was determined within the framework of the  dynamical spinodal,
i.e. the surface that limits the region where nuclear matter is
unstable with respect to density fluctuations as defined in \cite{Providencia-06}.
This method had been applied to the study of the fragmentation of a finite
nuclear system
within a self-consistent quantum approach in Refs.
\cite{colonna02,chomaz04}.  

Results in \cite{Fang16,Fang17} indicate
that the region defined by the $B=0$   spinodal section is not much
affected for finite $B \lesssim 10^{17}$G. However, besides this main
region of instability, other  unstable regions were identified at
larger densities. The same conclusion was drawn within a
thermodynamical spinodal approach \cite{Chen17,Fang17a,Chatterjee2018}. One expects that pasta  configurations will emerge
at these densities, and this is the main
objective of the present work: to investigate the possible  extension
of the nonhomogeneous matter region
and the existence of islands
of clusterized matter above the $B=0$ crust-core transition. 

In Refs.~\cite{Lima-13,Bao21},  the effect
of strong magnetic fields on the inner crust constitution has been studied using a Thomas-Fermi
approach. However, the magnetic field intensities considered were
generally quite high,  and the region
above the $B=0$ crust-core transition was not investigated.  According
to Ref.~\cite{Sengo20}, the neutral line of a
poloidal magnetic field may fall inside the extended nonhomogeneous
region. The existence of these regions may
be important to  avoid the development of magnetic field
instabilities \cite{Lander2020}.

 The crust-core transition density depends on
the properties of the nuclear equation of state, in particular, on its
symmetry energy, as discussed in several works
\cite{Xu2009,Vidana2009,Ducoin2010,Ducoin2011,Newton2013,Pais2016Vlasov,Li2020}. This has led us to
consider two models with the same isoscalar properties but different
isovector ones, in order to determine differences associated
 with the density dependence of the symmetry energy. In particular,
 {\it  ab-initio}
 chiral effective field theory calculations and several experimental
 nuclear properties seem to favor a small slope of the symetry energy
 at saturation, $L \lesssim 60$ MeV \cite{Lattimer2013}. However, if astrophysical
 obervations are also considered, larger slopes are predicted
 \cite{Oertel2017}, but still below 90 MeV. Recently, however, the measurement of the skin
 thickness of $^{208}$Pb \cite{PREX2} seems to indicate that  values
 of the slope $L$ in the interval 106 $\pm$ 37 MeV are expected \cite{Reed2021}, although not all
 studies estimate so large values, see \cite{Yue2021,Essick2021}. A
 large value of $L$ is also compatible with the estimate
 obtained from  charged pion spectra measured at high transverse
 momenta where $42<L<117$ MeV was obtained \cite{Estee2021}.

The structure of the paper is as follows: after the Introduction, we briefly review the formalism used to
determine the equation of state of magnetized matter in Sec.~\ref{eos}, and in Sec.~\ref{pasta}, clusterized matter is determined from the Gibbs conditions supplemented by  
Coulomb and finite size effects. In Sec.~\ref{sec:results}, the main results
are discussed, and in the last section, some conclusions are drawn.

\section{Magnetized nuclear matter \label{eos}}

\begin{figure*}[htp]
  \begin{tabular}{ccc}   
  \includegraphics[width=0.33\textwidth]{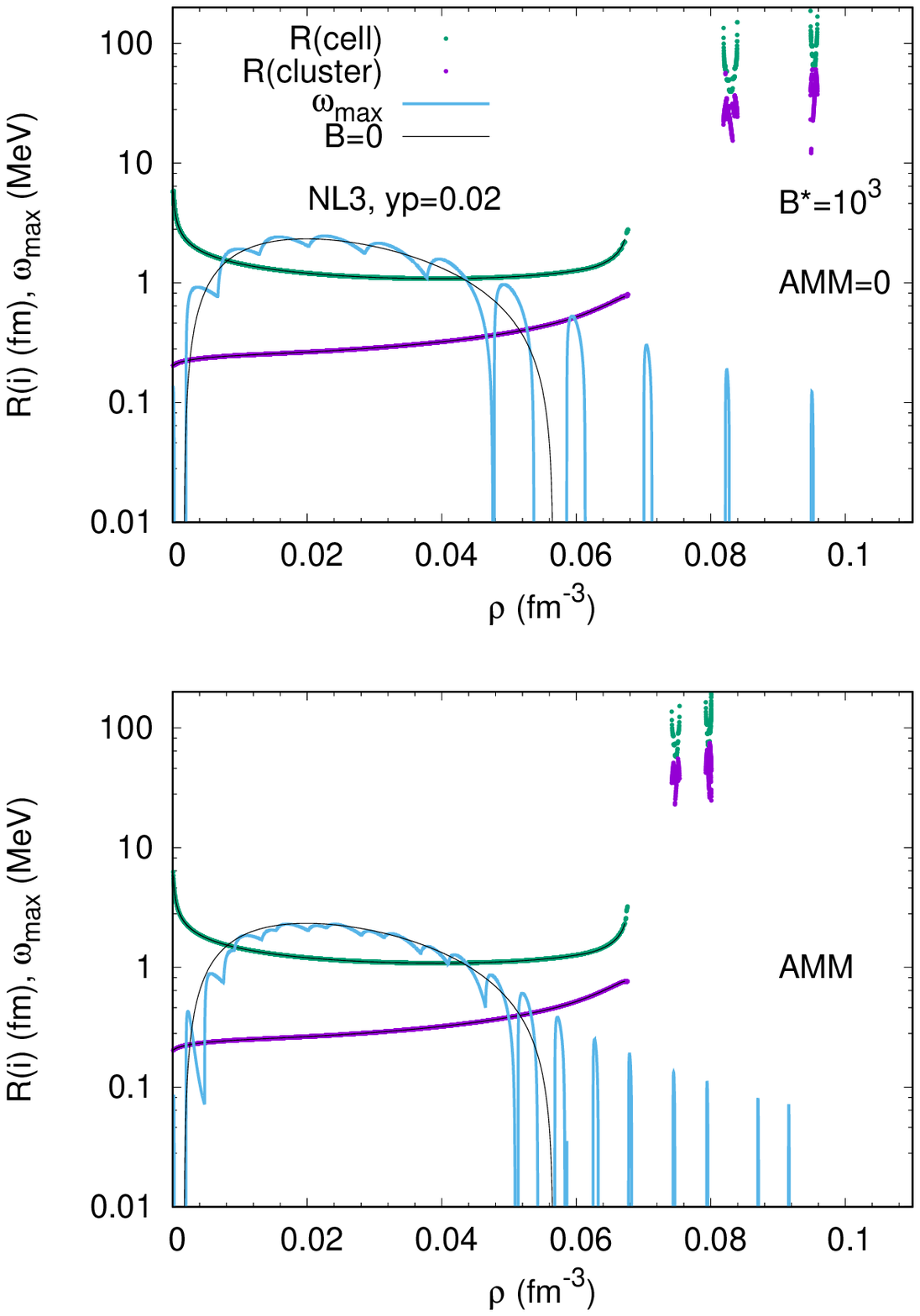} & 
   \includegraphics[width=0.33\textwidth]{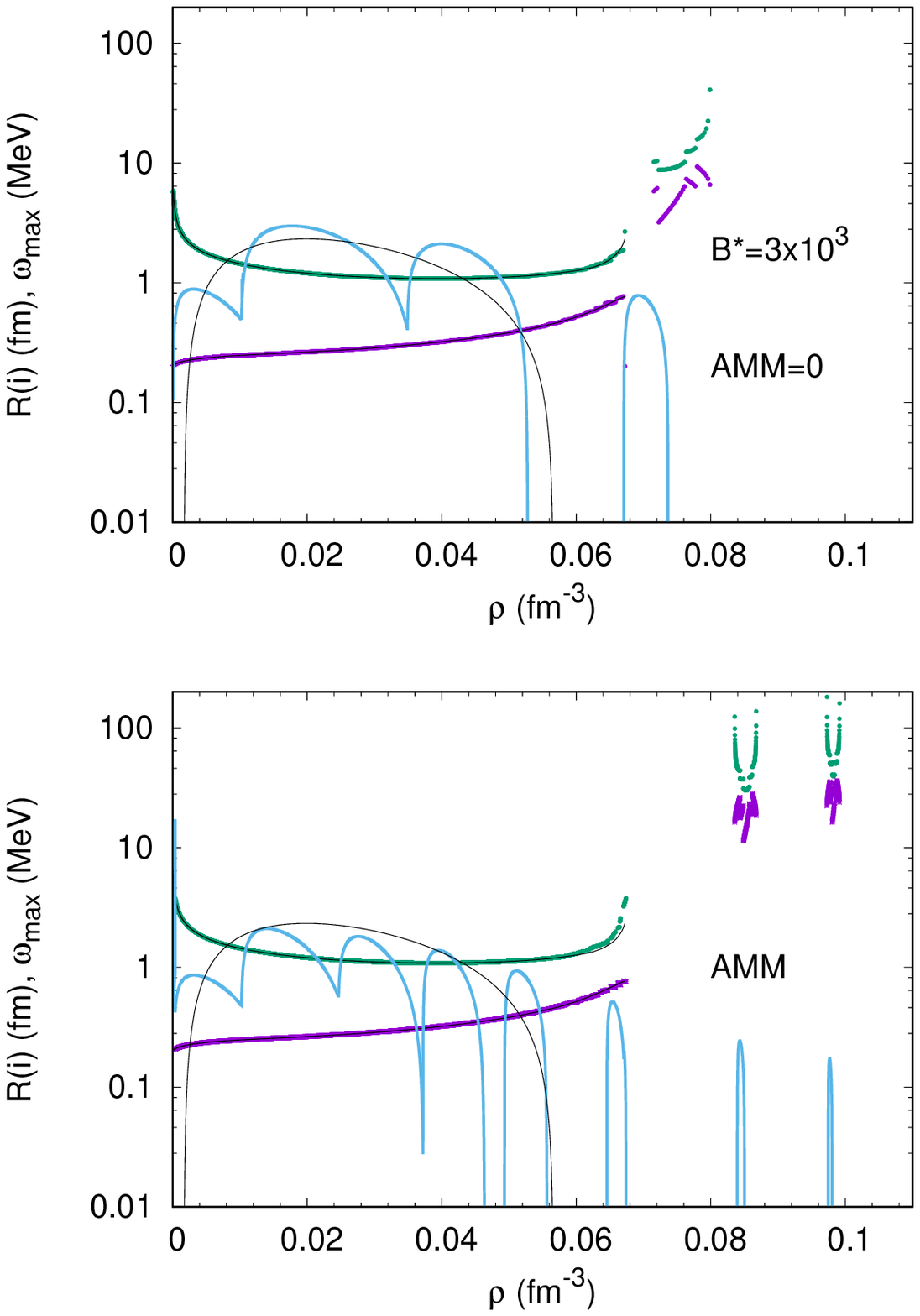} & 
  \includegraphics[width=0.33\textwidth]{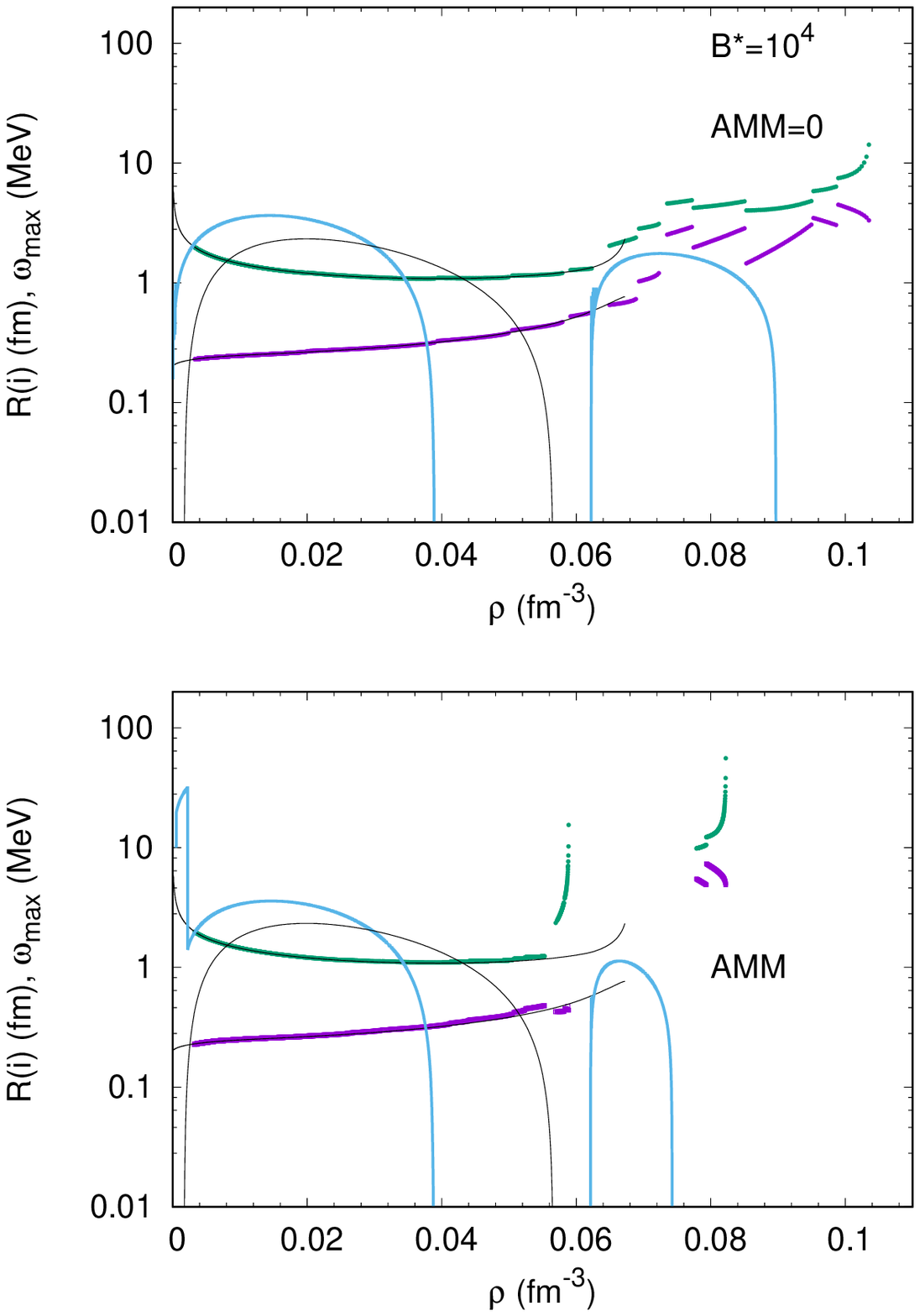} \\
    \end{tabular}
    \caption{Wigner-Seitz cell (thick green lines) and cluster radii
      (thick purple lines)  ($R_{WS}$ and $R_{D}$) and maximum growth rates
  ($|\omega|$) (thin black solid lines for $B=0$ and light blue solid lines for finite $B$)   for  $B= 4.4 \times 10^{16}$G (left), $B= 1.3 \times 10^{17}$G
  (middle), and $B= 4.4 \times 10^{17}$G  (right) determined for
  $y_p=0.02$ without (top) and with (bottom) AMM for the NL3 model.
 } 
\label{fig1}
\end{figure*}

In the present study we describe clusterized  stellar  matter   under
the effect of strong magnetic fields within
a nuclear relativistic mean-field  (RMF) approach
\cite{Broderick2000,Rabhi08}. In the following,  we  also discuss
the effect of the anomalous
magnetic moment (AMM) on the extension of the clusterized phase.

 In the RMF approach, nucleons with mass $M$ interact with and through
 the exchange of 
 three different mesonic fields, an
isoscalar-scalar field $\phi$ with mass $m_\sigma$ and coupling constant $g_\sigma$, an isoscalar-vector
field $V^{\mu}$ with mass $m_v$ and coupling constant $g_v$, and an isovector-vector field
$\mathbf b^{\mu}$ with mass $m_\rho$ and coupling constant $g_\rho$.  In order to describe
eletrically neutral matter, electrons will
also be considered explicitly.
The charged particles interact through the static electromagnetic field $A^{\mu}$,
$
 A^{\mu}=(0,0,Bx,0),
$
so that $\bf B$=$B$ $\hat{z}$ and $\nabla \cdot {\bf A}$=0.
In the present study, we consider that i) the
electromagnetic field is externally generated, ii)  field
configurations are frozen.

Our system is described by the Lagrangian density:% 
\begin{eqnarray}
{\cal L}=\sum_{i=p,n} {\cal L}_i + {\cal L}_e + {\cal L}_{A} + \cal L_\sigma + {\cal L}_\omega + {\cal L}_\rho + {\cal L}_{\omega\rho},
\end{eqnarray}
where ${\cal L}_i$ is the nucleon Lagrangian density, given by
\begin{eqnarray}
{\cal L}_i&=&\bar \psi_i\left[\gamma_\mu i D^\mu-M^*-\frac{1}{2}\mu_N\kappa_b\sigma_{\mu \nu} F^{\mu \nu}\right]\psi_i,
\end{eqnarray}
with
\begin{eqnarray}
M^*&=&M-g_\sigma\phi, \\
iD^\mu&=&i \partial^\mu-g_v V^\mu-
\frac{g_\rho}{2}\boldsymbol\tau \cdot \mathbf{b}^\mu - e A^\mu
\frac{1+\tau_3}{2}.
\end{eqnarray}
The electron Lagrangian density, ${\cal L}_e$ and the electromagnetic
term, ${\cal L}_{A}$, are defined  by
\begin{eqnarray}
{\cal L}_e&=&\bar \psi_e\left[\gamma_\mu\left(i\partial^\mu + e A^\mu\right)-m_e\right]\psi_e, \\
  {\cal L}_{A}&=&-\frac{1}{4}F_{\mu\nu}F^{\mu\nu}.
                  \label{lagr}
\end{eqnarray}
We consider $c=\hbar=$1, the electromagnetic coupling constant is $e=\sqrt{4\pi/137}$, and
$\tau_{3}=\pm 1$ is the
isospin projection for  protons and neutrons,
respectively. $M^*$ is the nucleon effective mass, and $m_e$ the electron mass. The inclusion of the AMM is undertaken via the coupling of the nucleons to the electromagnetic
field tensor, $F_{\mu\nu}=\partial_\mu A_\nu-\partial_\nu A_\mu$, with $\sigma_{\mu\nu}=\frac{i}{2}\left[\gamma_{\mu},
  \gamma_{\nu}\right] $, and strength $\kappa_{b}$, with
$\kappa_{n}=-1.91315$ for the neutron, $\kappa_{p}=1.79285$ for the
proton, and $\mu_N$ the nuclear magneton. It was discussed in
\cite{Duncan-00}, that the  contribution of the AMM of
electrons is negligible and, therefore, it will not be
included.  Notice that  medium effects may affect the AMM,
 as it was shown, for instance, in \cite{Frank1996}. In the present study, we consider
 the vacuum values.

The
mesonic  Lagrangians are given by
\begin{eqnarray}
{\cal L}_\sigma&=&\frac{1}{2}\left(\partial_\mu\phi\partial^\mu\phi-m_\sigma^2 \phi^2 - \frac{1}{3}\kappa \phi^3 -\frac{1}{12}\lambda\phi^4\right),\nonumber\\
{\cal L}_\omega&=&-\frac{1}{4}\Omega_{\mu\nu}\Omega^{\mu\nu}+\frac{1}{2}
m_v^2 V_\mu V^\mu, \nonumber \\
{\cal L}_\rho&=&-\frac{1}{4}\mathbf B_{\mu\nu}\cdot\mathbf B^{\mu\nu}+\frac{1}{2}
m_\rho^2 \mathbf b_\mu\cdot \mathbf b^\mu,
\end{eqnarray}
where
$\Omega_{\mu\nu}=\partial_\mu V_\nu-\partial_\nu V_\mu$, and $\mathbf B_{\mu\nu}=\partial_\mu\mathbf b_\nu-\partial_\nu \mathbf b_\mu
- g_\rho (\mathbf b_\mu \times \mathbf b_\nu)$. $\kappa$ and $\lambda$ are the third- and fourth-order parameters of the scalar field.  

In order to discuss the effect of the density dependence of the
symmetry energy we consider two models, NL3 and NL3$\omega\rho$, that have the
same isoscalar properties but different isovector properties, as it was
considered in \cite{Fang16}.
Therefore, we also include in the Lagrangian density the nonlinear term, ${\cal L}_{\omega\rho}$, that mixes the $\omega$ and $\rho$ mesons
\begin{eqnarray}
\mathcal{L}_{\omega \rho } &=& \Lambda_v g_v^2 g_\rho^2 V_{\mu }V^{\mu }
\mathbf{b}_{\mu }\cdot \mathbf{b}^{\mu }.
\end{eqnarray}
This term allows to have a softer symmetry energy than the one of the
original NL3 model  \cite{nl3}. This mechanism of modelling the density dependence of the symmetry energy using a non-linear term that mixes the $\omega$ and $\rho$ mesons was discussed in \cite{Horowitz2001PRL}.

\section{The pasta phases\label{pasta}}

    In the present study we consider the coexisting phases (CP) method to
      describe  the pasta phases as discussed in \cite{Maruyama2005,Avancini-08,Avancini-10,Bao-14,Pais-15}.  A
      self-consistent description within a Thomas-Fermi approximation has
      already been applied to the description of magnetized matter \cite{Lima-13,Bao21}. We will consider such a calculation in a future study since numerical convergence for magnetic
      field intensities $\lesssim 5\times 10^{17}$G is faster within the CP approach we are presently considering.
 In this approach, the free energy density is minimized considering a mixed phase
      of dense clusters with different geometrical configurations in a
      background gas of nucleons and electrons. Finite size effects, as the ones from the cluster
      surface tension and the Coulomb interaction, are included after
      the minimization of the free energy.

The free energy density minimum is
determined by imposing the Gibbs conditions on the pressure $P$, and proton
and neutron chemical potentials, respectively $\mu_p$ and $\mu_n$,
 at
the transition from phase I to phase II \cite{Maruyama2005,Avancini-08,Avancini-10,Bao-14,Pais-15},
\begin{equation}
P^I({E^i_{F}}^I,{M^*}^{I})=P^{II}({E^i_{F}}^{II},{M^*}^{II}),
\quad i=p,n\label{gibbs1}
\end{equation}
\begin{equation}
\mu_i^I=\mu_i^{II}, \quad i=p,n . \label{gibbs2}
\end{equation}

 In the above equations, the chemical potentials $\mu_i$ are given by
\begin{eqnarray}
\mu_p&=& E^p_{F}+g_{\omega} V^{0}+\frac{1}{2}g_{\rho} b^{0}, \label{mup}\\
\mu_{n}&=&E^n_{F}+g_{\omega} V^{0}-\frac{1}{2}g_{\rho} b^{0}\label{mun},
\end{eqnarray}
with $E^i_{F}$ being the effective chemical potentials.

 The Gibbs equilibrium conditions are supplemented by the equations that define the nucleon effective masses in both phases,
\begin{equation}
m_\sigma^2 \phi_0^I + \frac{\kappa}{2} {\phi_0^2}^I 
+\frac{\lambda}{6}{\phi_0^3}^I = g_\sigma \rho_s^I,\label{gibbs3}
\end{equation}
\begin{equation}
m_\sigma^2 \phi_0^{II} + \frac{\kappa}{2} {\phi_0^2}^{II} 
+\frac{\lambda}{6}{\phi_0^3}^{II} = g_\sigma \rho_s^{II},\label{gibbs4}
\end{equation}
 and by the equation that fixes the global proton fraction
\begin{equation}
f\rho_p^I + (1-f) \rho_p^{II} = Y_p \rho. \label{gibbs7}
\end{equation}

In the above equations,  I refers to the liquid (cluster) phase and
II to the gas phase, and $f$ is the volume fraction of 
phase I: 
\begin{equation}
f= \frac{\rho -\rho^{II}}{\rho^I-\rho^{II}}.
\end{equation}

 $\rho_s=\rho_{s,p}+\rho_{s,n}$ is the total scalar density
with $\rho_{s,p}, \, \rho_{s,n}$ the proton and
neutron scalar densities, respectively, and $\rho=\rho_p+\rho_n$ is the total nucleonic density, with $\rho_{p}, \, \rho_{n}$ the proton and
neutron densities, respectively. They are going to be defined next. $Y_p$ is the global proton fraction, $Y_p=\rho_p/\rho$. We also consider that the density of electrons $\rho_e$ is uniform in the
cell and $\rho_e=Y_p
\rho$.

The scalar and vector proton and neutron  densities  in the above equations are defined by \cite{Broderick2000,PerezGarcia2011}
\begin{eqnarray}
\rho_{s,p}&=&\frac{q_{p}B M^*}{2\pi^{2}}\sum_{\nu=0}^{\nu_{\rm max}}\sum_{\varsigma}\frac{\sqrt{M^{* 2}+2\nu 
q_{p}B}-{\varsigma}\mu_{N}\kappa_{p}B}{\sqrt{M^{* 2}+2\nu q_{p}B}} \nonumber\\ &&\ln\left|\frac{k^{p}_{F,\nu, {\varsigma}}+E^{p}_{F}}
                                                                            {\sqrt{M^{*
                                                                        2}+2\nu
                                                                        q_{p}B}-{\varsigma}\mu_{N}\kappa_{p}B}
                                                                        \right|,
                                                                        \label{rhosp} 
\end{eqnarray}
\begin{eqnarray}
\rho_{s,n}&=&\frac{M^{*}}{4\pi^{2}}\sum_{{\varsigma}} \left[E^ {n}_{F}k^{n}_{F, {\varsigma}}-\bar{m}^{2}_{n}\ln\left|
\frac{k^{n}_{F, {\varsigma}}+E^{n}_{F}}{\bar{m}_{n}} \right|\right], \label{rhosn} \\
\rho_{p}&=&\frac{q_{p}B}{2\pi^{2}}\sum_{\nu=0}^{\nu_{\rm
      max}}\sum_{{\varsigma}}k^{p}_{F,\nu,\varsigma}, \label{rhop}  \\ 
\rho_{n}&=&\frac{1}{2\pi^{2}}\sum_{{\varsigma}}\left[ \frac{1}{3}\left(k^{n}_{F, {\varsigma}}\right) ^{3}-\frac{1}
            {2}{\varsigma}\mu_{N}\kappa_{n}B \right.\times \nonumber\\
  && \left.\left(\bar{m}_{n}k^{n}_{F, {\varsigma}}+E^{n 2}_{F}\left(\arcsin\left( \frac{\bar{m}_{n}}
{E^{n}_{F}}\right) -\frac{\pi}{2} \right)  \right) \right]\, .
\label{rhon}
\end{eqnarray}

 In the above equations, $\kappa_i,\, i=p,n$ are the anomalous
magnetic moment strengths, previously defined after Eq. (\ref{lagr}).
$k^{p}_{F, \nu, {\varsigma}}$, $ k^{n}_{F, {\varsigma}}$ are the Fermi momenta of protons and neutrons
related to the proton and neutron effective Fermi energies
\begin{eqnarray}
k^{p 2}_{F,\nu, {\varsigma}}&=&{E^p_{F}}^{2}-\left[\sqrt{M^{* 2}+2\nu q_{p}B}-{\varsigma}\mu_{N}\kappa_{p}B\right] ^{2}, \label{kfp}\\
k^{n 2}_{F, {\varsigma}}&=&{E^n_{F}}^{2}-\bar{m}^{2}_{n}, \label{kfn}
\end{eqnarray}
 with $\bar{m}_{n}=M^{*}-{\varsigma}\mu_{N}\kappa_{n}B$.
 $\nu=n+\frac{1}{2}-\frac{{\varsigma}}{2}=0, 1, 2, \ldots$ enumerates the quantized  Landau levels for  protons with charge $q_p=1$, and ${\varsigma}$ is the spin quantum number, with $+1$ for spin up and $-1$ for spin down.

The solution of the equations (\ref{gibbs1}-\ref{gibbs7}) defines the total
energy density of the nonhomogeneous matter,
\begin{equation}
{\cal E}= f {\cal E}^I + (1-f) {\cal E}^{II} + {\cal E}_e +
{\cal E}_{surf} + {\cal E}_{Coul}, 
\label{totener}
\end{equation}
where ${\cal E}^{I,II} $ refer to the bulk free energy density of
phase $I$ and $II$,  $ {\cal E}_e$ is the free energy
density of an homogeneous eletron gas, and the terms ${\cal E}_{surf}$
and ${\cal E}_{Coul}$ were added to take into account, respectively,  the surface
energy of the clusters and the Coulomb interaction. From the 
minimization of the sum  ${\cal E}_{surf} + {\cal E}_{Coul}$ with respect
to the size of the cluster, the following relation is obtained
\cite{Maruyama2005}
${\cal E}_{surf} = 2 {\cal E}_{Coul}.$
The Coulomb energy density is defined by \cite{Ravenhall-83,Hashimoto-84,Maruyama2005}
\begin{equation}
{\cal E}_{Coul}=\frac{2 F}{4^{2/3}}(e^2 \pi \Phi)^{1/3} 
\left(\sigma D (\rho_p^I-\rho_p^{II})\right)^{2/3},
\end{equation}
where $F=f$ is used in the droplet/rod/slab phase and  and $F=1-f$ for
tubes and bubbles, 
 $\sigma$ is the surface energy coefficient and 
$D$ is the dimension of the system. For droplets, rods and slabs, \cite{Ravenhall-83,Hashimoto-84}
\begin{equation}
\Phi=\begin{cases}
\left(\frac{2-D F^{1-2/D}}{D-2}+F \right) \frac{1}{D+2}, \quad D=1,3;\\
 \frac{F-1-ln(F)}{D+2}, \quad D=2. \end{cases}
\end{equation}
In the present study, we define the surface energy as in Ref.~\cite{pasta_alpha}, where a parametrization determined within a Thomas-Fermi approach was obtained in terms of the global proton fraction and the temperature,
$\sigma (x,T)$, with $x=(1-2 Y_p)^2$. Each
cluster is considered to be in the center of a charge neutral 
Wigner-Seitz  (WS) cell \cite{Maruyama2005,Shen98}.  
The WS cell may have different geometries, and for
simplification they are fixed to a sphere, a cylinder, or a slab, and its volume is equal to the one of the unit
BCC cell. The dimensions of the WS cell ($R_{WS}$)  and of the cluster
($R_D$) are
determined from the  minimization of the free energy with respect to the
cluster dimension  $R_D$ (the radius of the
droplet or rod and the thickness of the slab). The following
expressions define $R_D$ and $R_{WS}$, see \cite{Ravenhall-83,Maruyama2005}:
\begin{equation}
R_D=\left( \frac{\sigma D}{e^2 (\rho_p^I
- \rho_p^{II})^2 \Phi} \right)^{1/3},
\end{equation}
\begin{equation}
R_{WS}=\frac{R_D}{F^{1/D}}. 
\end{equation}

\section{Results}
\label{sec:results}

In the following, the nonhomogeneous matter is calculated according to
the formalism presented in the previous section.  We consider two models that
have the same isoscalar properties, but differ on the isovector ones,
NL3 \cite{nl3} and NL3$\omega\rho$ \cite{Pais2016Vlasov}. The symmetry
energy  slope at saturation  is equal to 118 MeV for NL3 and 55 MeV
for NL3$\omega\rho$.
 Recently, an estimation of $L$ obtained from the
 measurement of the $^{208}$Pb neutron skin thickness \cite{PREX2}  predicts
 $106\pm37$ MeV \cite{Reed2021}.
However, other studies based on experimental
measurements and ab-initio calculations indicate a much smaller slope, 
 $L= 51\pm11$ MeV \cite{Lattimer2013}, or taking into account also
 astrophysical observations, $L= 58.7\pm28.1$ MeV
 \cite{Oertel2017}.
 The two values of $L$ we consider represent both scenarios.
 
It was shown in Ref.~\cite{Grill2012} that in the lower layers of the inner
crust, close to the crust-core transition, the proton fraction does not
change much and takes a value close to the one at the transition. For
the above models, this corresponds to $y_p=0.02$ for NL3 and
$y_p=0.035$ for NL3$\omega\rho$.

In the following, we will study the formation of pasta phases for
those proton fractions as an exploratory investigation. 

\subsection{Pasta structures at supra crust-core transition density}

We first consider the NL3 model.  In Fig.~\ref{fig1},  the size of
the WS cell $R_{WS}$ (green
points) and the cluster size $R_D$ (purple points) are plotted together with the growth
rate determined within a dynamical spinodal calculation (light blue
lines), for magnetic fields with intensities  0.44, 1.3 and 4.4
$\times 10^{17}$G. This corresponds to $B^*= 10^3,\, 3\times 10^3$ and
$10^4$, respectively, because $B^*=B/B_{ce}$, with $B_{ce}=4.41\times
10^{13}$ G the critical electron magnetic field.
The growth rates have been defined in \cite{Fang17}, see
  Eq. (36),  and are the
  solutions of the dispersion relation obtained considering
  fluctuations of the proton, neutron and electron  densities. Inside
  the dynamical spinodal, the solutions of the dispersion relation are
  imaginary frequencies characteristic of unstable modes. We define
  the growth rates as the modulus of these solutions. The magnitude of
  the growth rates depends on the wavelength of the mode, and, for each density
we consider the largest growth rate as the one that drives the system. The respective wavelength can be
considered an estimation of the size of the clusters formed \cite{Ducoin2007}.
Results obtained 
without (with)   the anomalous magnetic moment of the nucleons are
shown in the top (bottom) panels. 
The $B=0$ results are represented by thin black lines.

\begin{figure}[htp]
  \includegraphics[width=0.95\linewidth]{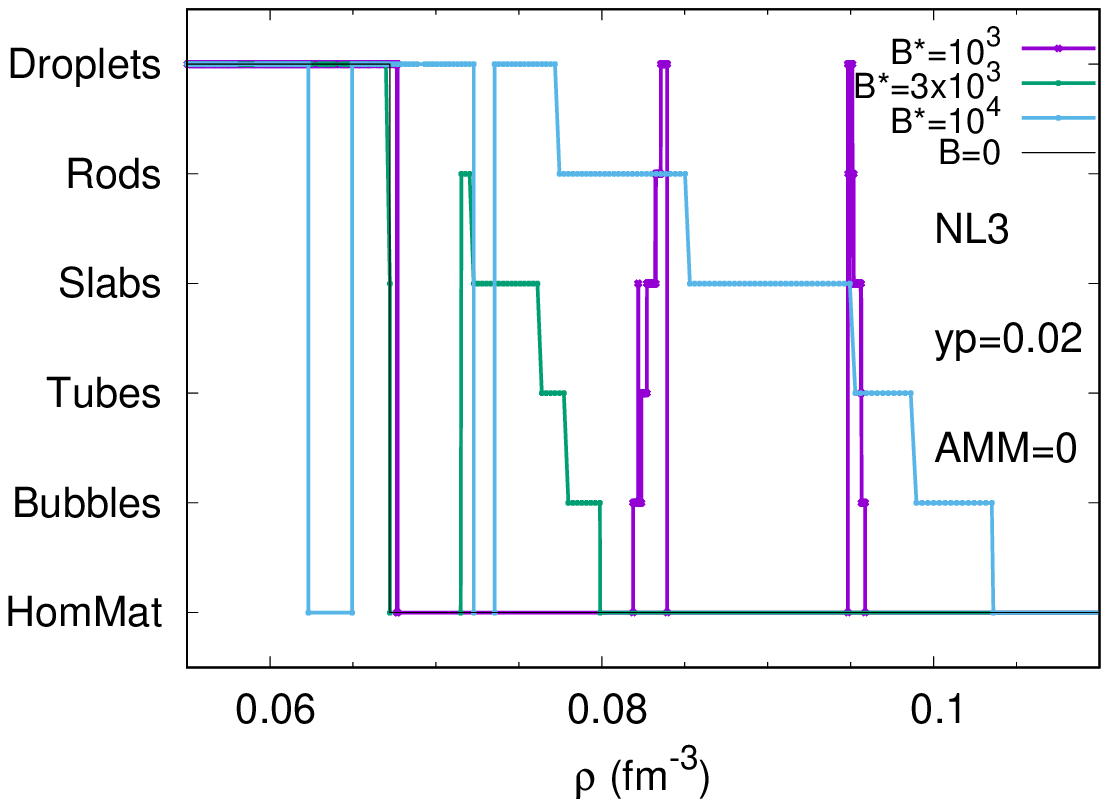} \\
  \includegraphics[width=0.95\linewidth]{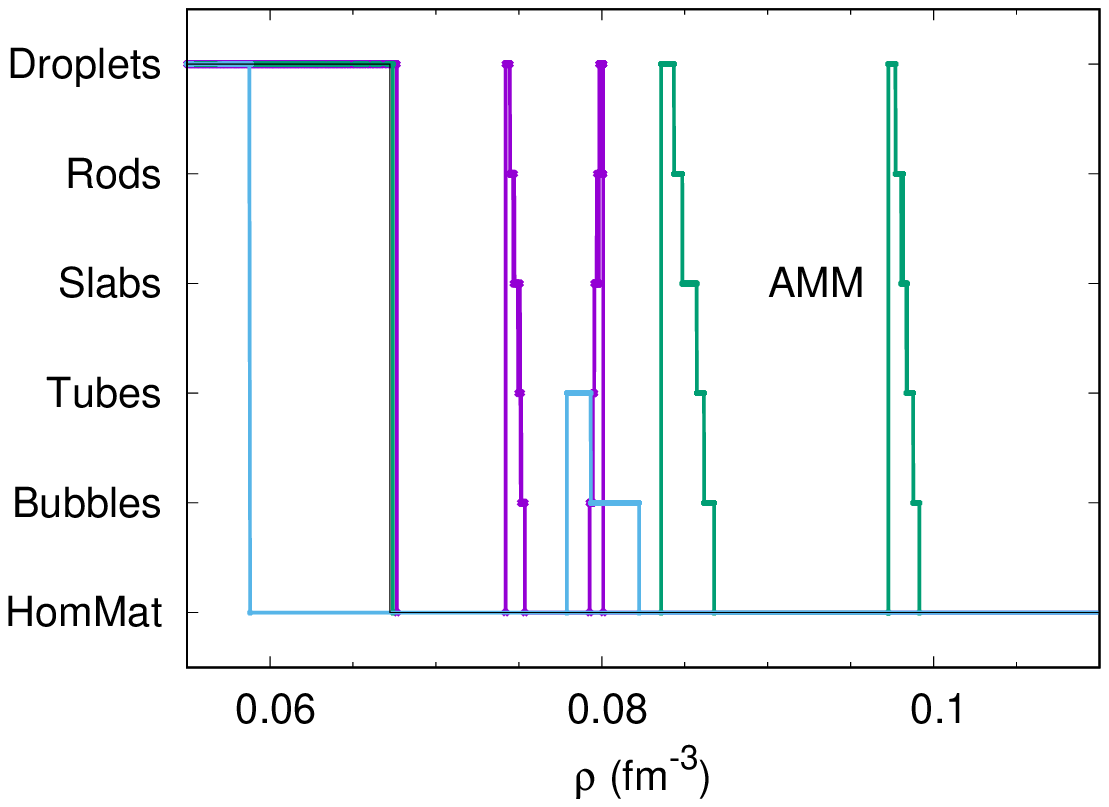} \\
  \caption{Geometries of the clusters obtained with  yp=0.02, without
    (top) and with (bottom)
    AMM within the NL3 model for
    several magnetic field intensities.} 
\label{shapesNL3}
\end{figure}

\begin{figure*}[htp]
  \begin{tabular}{ccc}
    \includegraphics[width=0.33\textwidth]{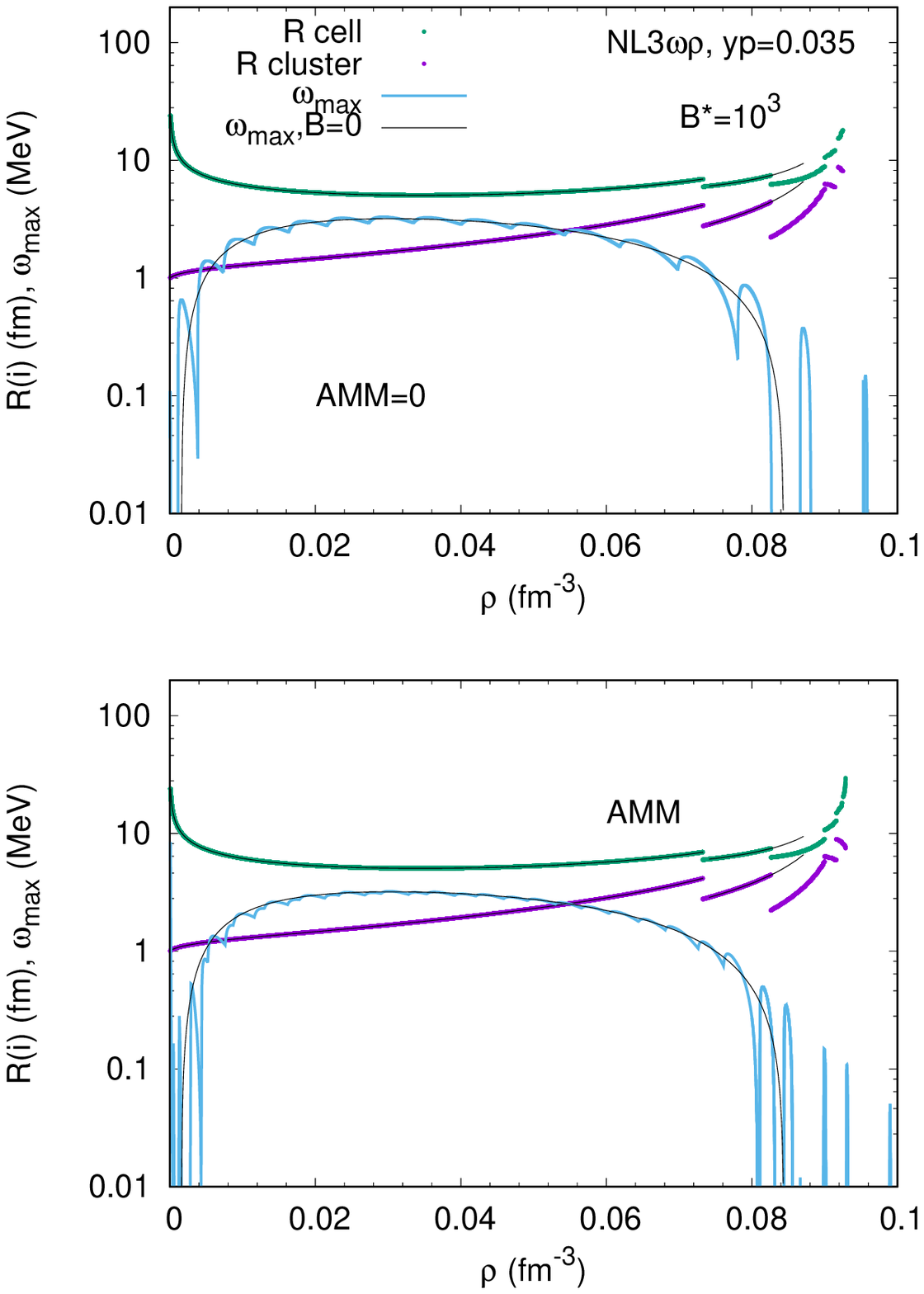} &                                                                       
   \includegraphics[width=0.33\textwidth]{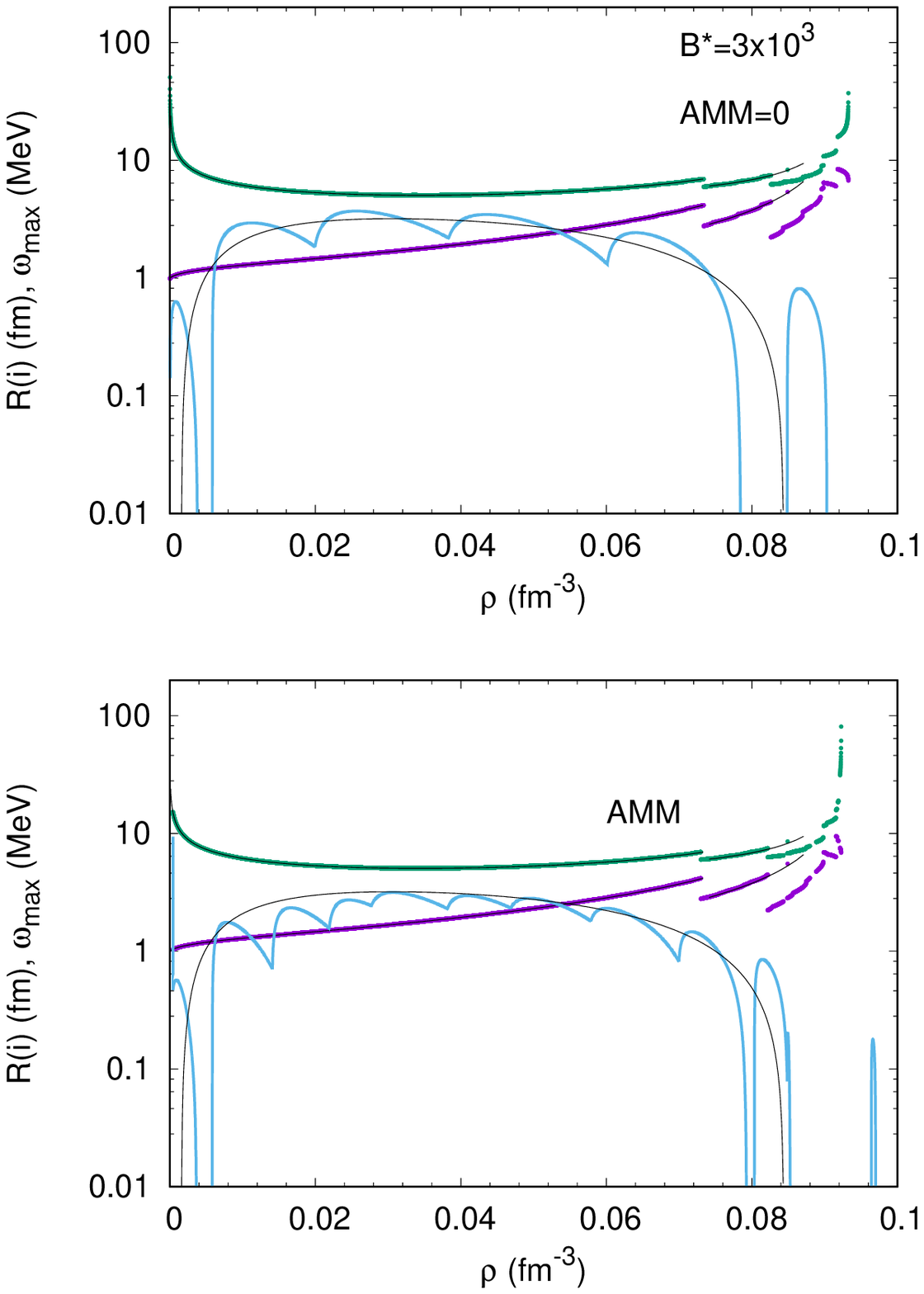} &
  \includegraphics[width=0.33\textwidth]{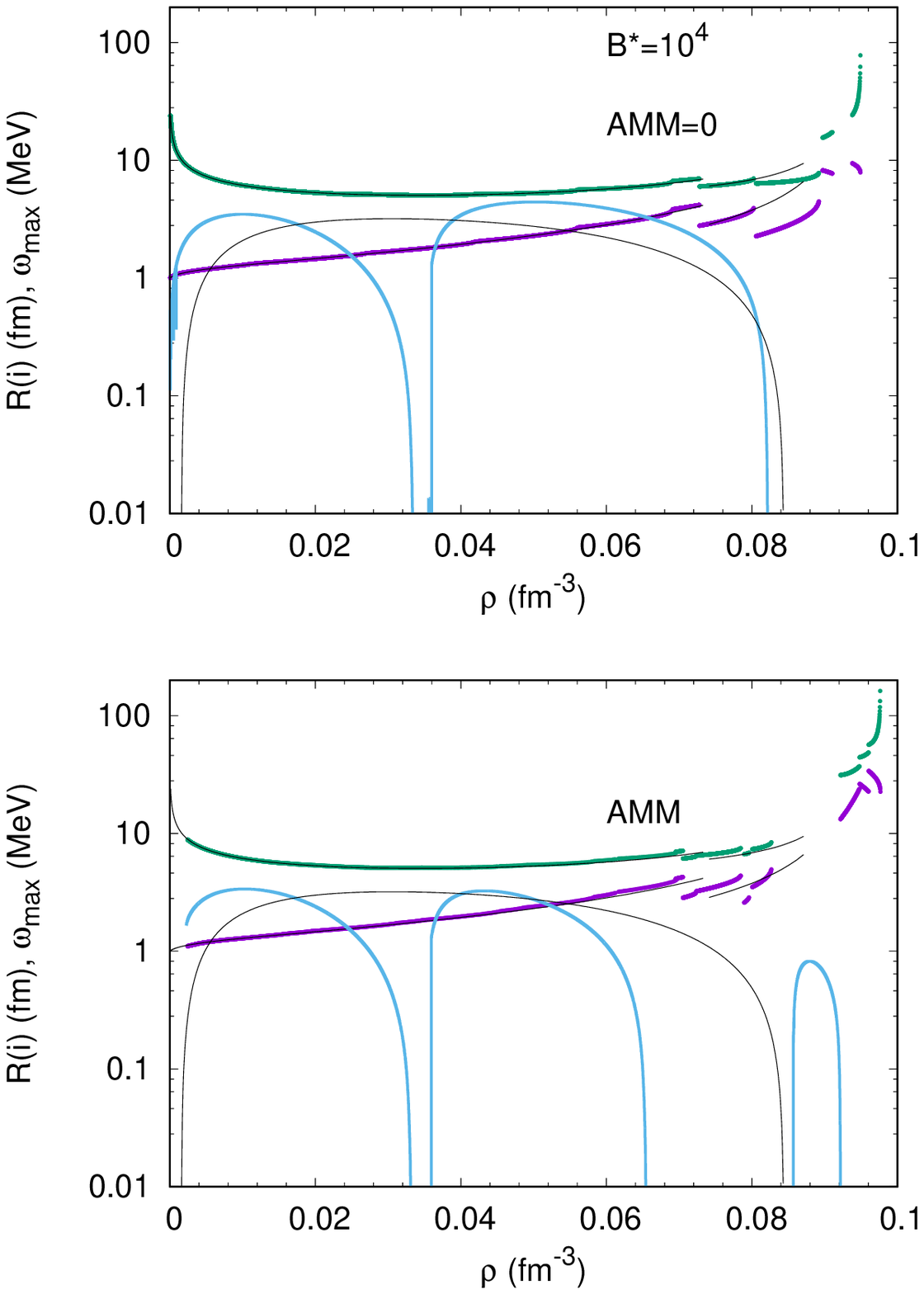} \\ 
  \end{tabular}
\caption{Wigner-Seitz cell (thick green curves) and cluster radii
  (purple thick lines) and maximum growth rates (thin black solid lines for $B=0$ and light blue solid lines for finite $B$) for  $B=4.4\times 10^{16}$G ($B^*= 10^{3}$, left),
  $B=1.3\times 10^{17}$G    ($B^*=3\times 10^{3}$, middle) and
   $B=4.4\times 10^{17}$G ($B^*= 10^{4}$, right), with (bottom) and without (top) AMM for the
  NL3$\omega\rho$ model. The proton fraction is fixed to 0.035.}
\label{nl3wr}
\end{figure*}

 Let us start by analysing the top panels. For the two weaker field intensities
represented, the low density region, i.e. the region that is also found in the $B=0$ calculations, does not seem to be sensitive to
the magnetic field. However, a new feature is present: a region of clusterized matter appears 
at larger densities, of the order of  0.08-0.09
fm$^{-3}$.
 This region is coincident with a region of instability identified
within the dynamical spinodal formalism.
For the magnetic  field intensity 4.4
$\times 10^{17}$G, the clusterized region extends until a density
above 0.1 fm$^{-3}$. 

In the following, we analyse the effect of including the proton and
  neutron AMM.  Taking as reference the energy
  $\epsilon=|\kappa_p+\kappa_n|B \approx 2\times 10^{-5}B^*$ MeV,
it is clear that the effect of the AMM is only strong for very strong
fields, i.e $B^*\gtrsim 10^5$. However, its
effect is seen as soon as the energy difference between different free
energy configurations is of the order of $\epsilon$.  The AMM term
gives rise to proton and neutron polarization and, therefore, 
reduces the softening  effect of the Landau
quantization on the EoS. 
  
Looking at the bottom panels, the introduction of the AMM in the calculation does
not change much the extension of the low density clusterized region with respect to the previous
situation, but the second region of clusterized matter decreases considerably, for the two highest fields considered. In particular, for the strongest field
considered, the inclusion of AMM in the calculation results in
  the division of the clusterized region into two disconnected
  clusterized regions separated by homogeneous matter.
For this same field, it is also seen that
the upper bound of the low density clusterized region occurs at smaller
densities, when compared to the $B=0$ field case \cite{Avancini-10}.
This is in agreement with the results presented in
\cite{Bao21}, where it is shown that the crust-core transition density
decreases for $B=10^{18}$G, when AMM is taken into account. In our
calculation, this decrease is only observed for  $B=4.4\times10^{17}$G:
the weaker fields do not affect this boundary. Another interesting
feature is the fact that  for  the weaker fields considered, more
  than two
disconnected regions of clusterized matter are obtained.

In Fig.~\ref{shapesNL3}, we identify the geometry of the clusters
obtained within the model NL3 without (top) and with  (bottom)
AMM. The same
magnetic field intensities as before are considered. Some conclusions can
 immediately be drawn: i) the low density behavior is very similar for the
three values of $B$. In this region, only droplets are found,
consistent with the $B=0$ case; ii)
 a second or more regions
of nonhomogeneous matter appear. In the AMM case, and for the two
lowest fields considered, two separate regions are found. The same
occurs if no AMM is included for the lowest field.
In these new regions of clusterized matter, all different geometries may
occur  with an extension and limiting densities very sensitive to
the magnetic field. As already discussed, the inclusion of the AMM has also
non-negligible 
effects. For the two highest fields considered, the supra-saturation nonhomogeneous region
suffers a shift  that depends on the magnetic field intensity
considered, and its density extension decreases. 

The strong dependence of the structure of
 these
other regions on the
magnetic field intensity may have direct effects on the
evolution of the magnetar. In fact, the decay of the magnetic field
will have consequences on the equilibrium structure of the neutron
star and originate internal stresses which may give rise to the yield or
the fracture of the  lattice.

In the following, we consider the calculation of the nonhomogeneous
matter within the NL3$\omega\rho$ model, a model that predicts a much
softer density dependence of the symmetry energy. In this case, the $B=0$
calculation predicts the existence of droplets, rods and slabs,  and
not only droplets. This behavior had
been predicted in earlier studies \cite{Oyamatsu2007}.

\begin{figure*}[htp]
  \begin{tabular}{ccc}
  \includegraphics[width=0.3\textwidth]{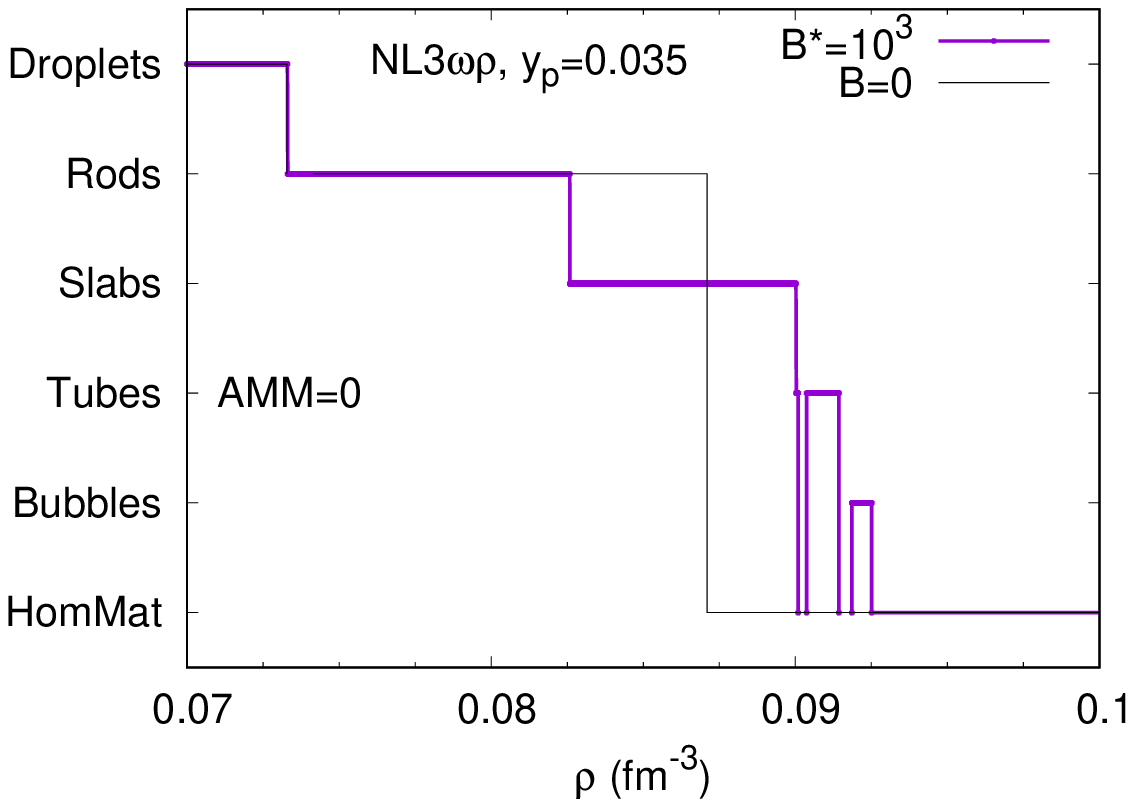} &  \includegraphics[width=0.3\textwidth]{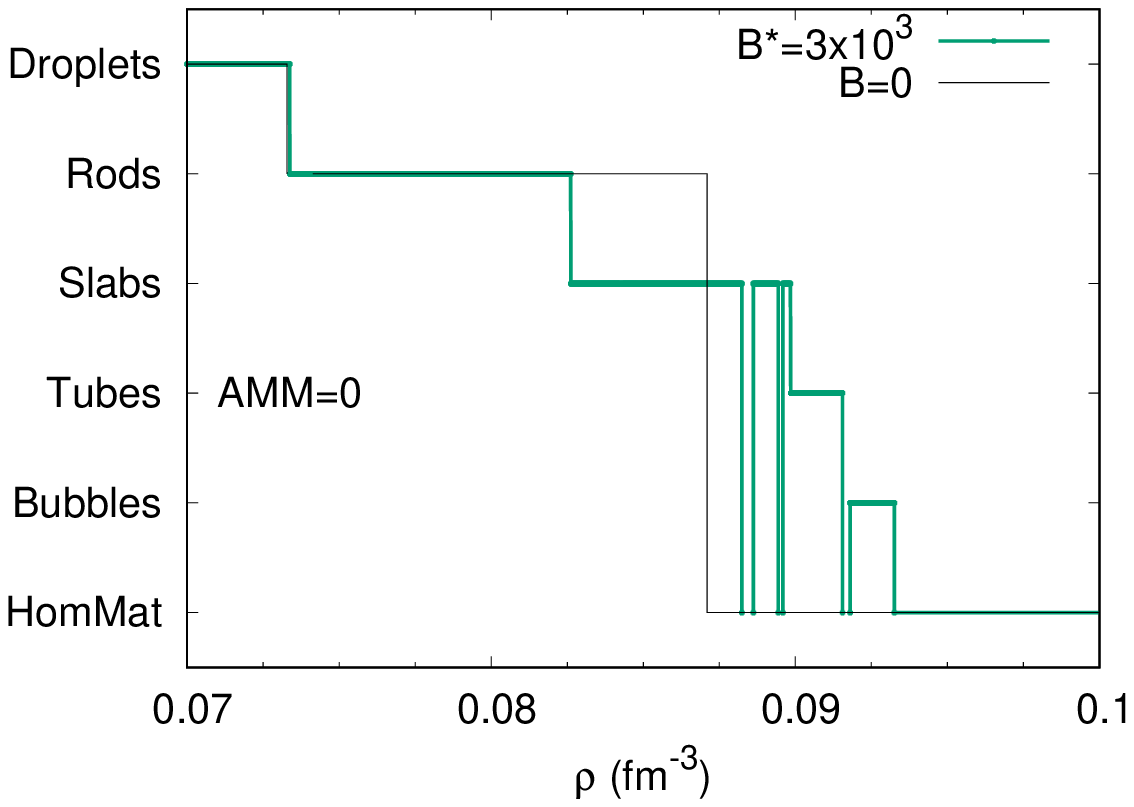} &  \includegraphics[width=0.3\textwidth]{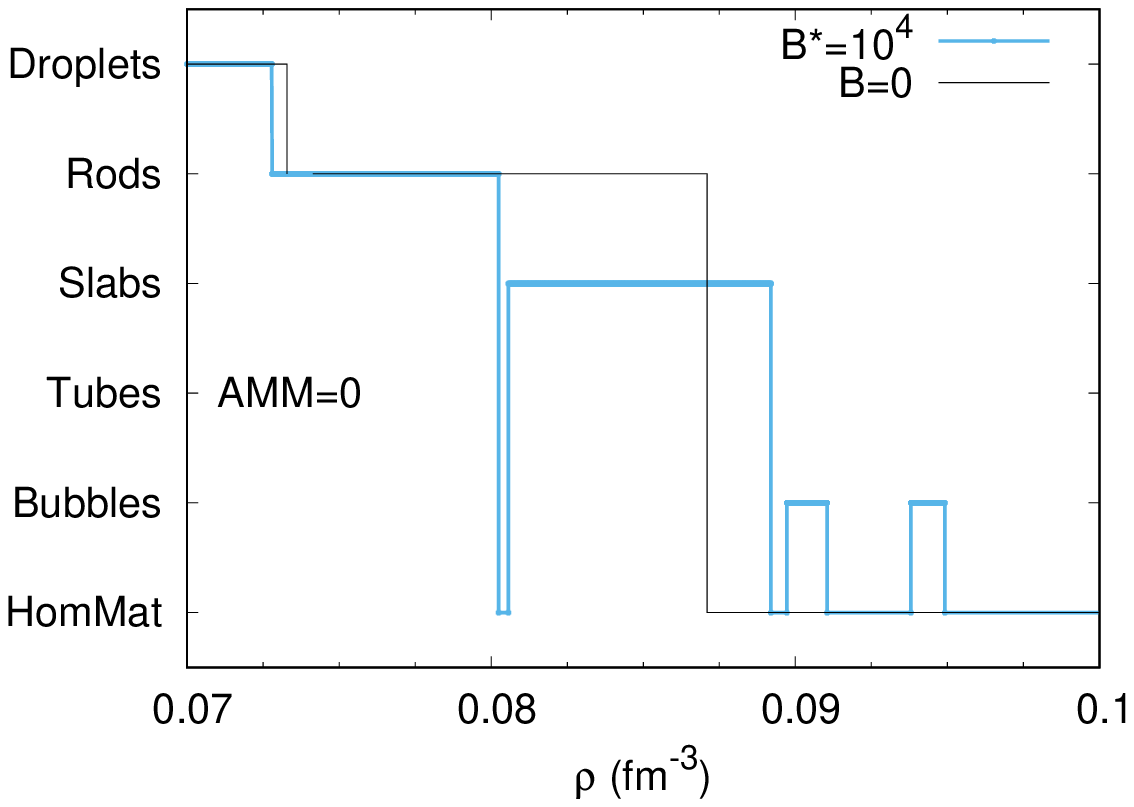} \\
  \includegraphics[width=0.3\textwidth]{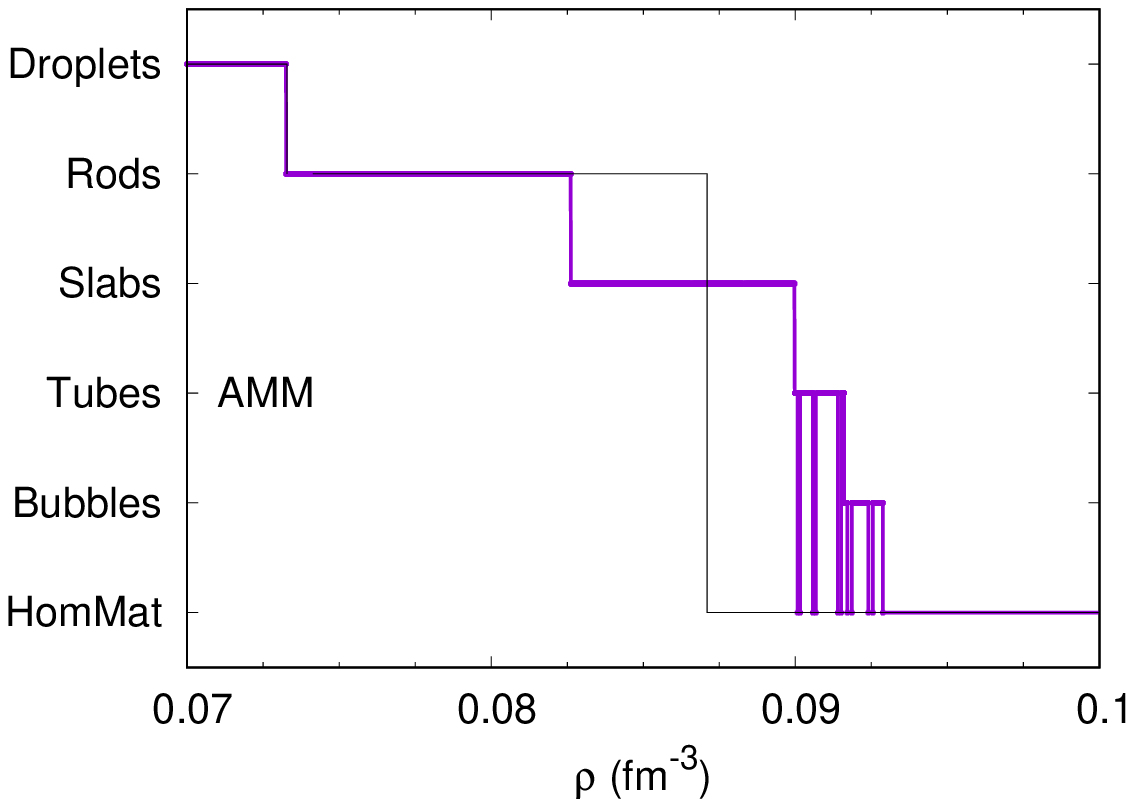} & \includegraphics[width=0.3\textwidth]{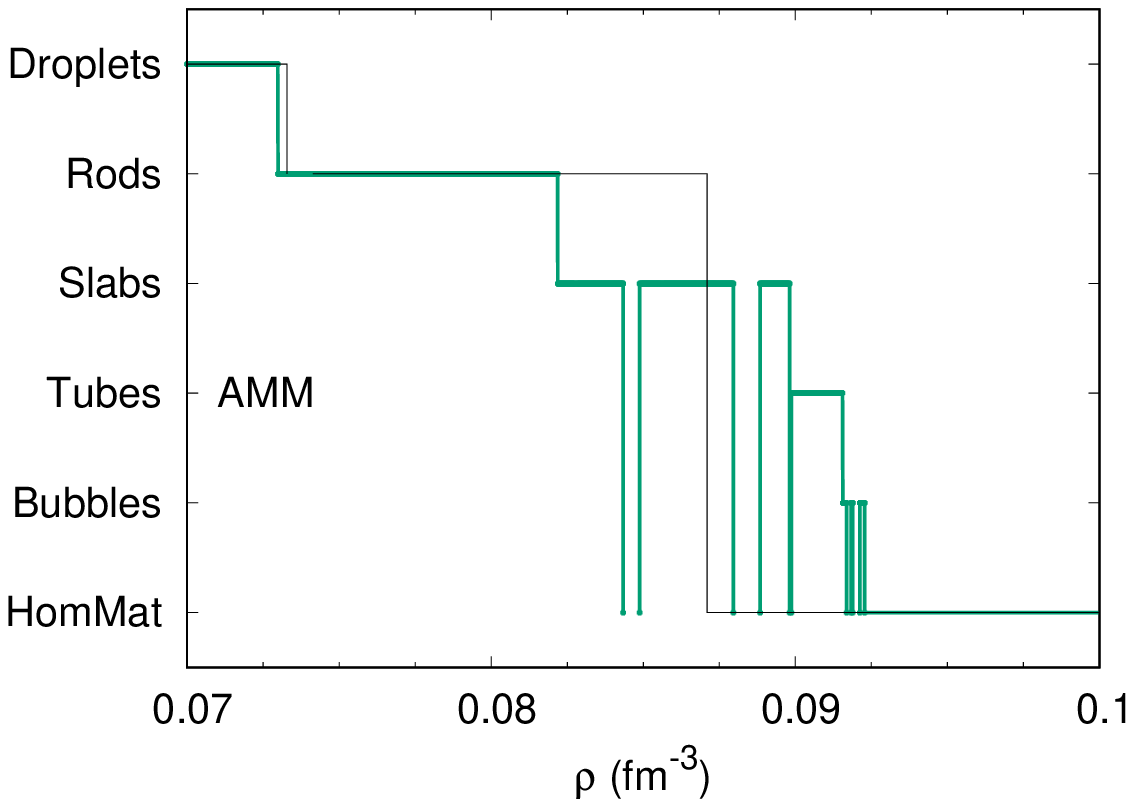} & \includegraphics[width=0.3\textwidth]{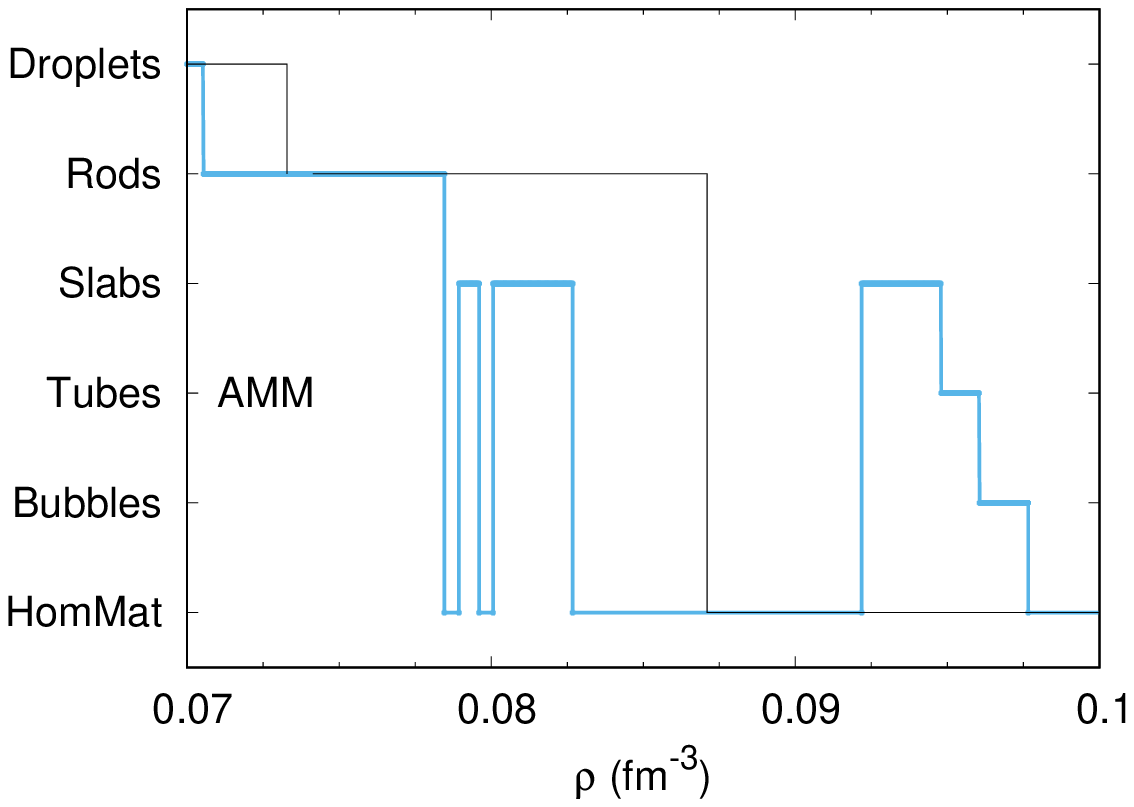} 
   \end{tabular}
\caption{Geometries of the clusters obtained with  yp=0.035, without
    (top) and with (bottom)
    AMM within the NL3$\omega\rho$ model for
    the magnetic field intensities  $B^*=10^3$ (left), $B^*=3\times 10^3$ (middle) and $B^*=10^4$ (right). The thin black lines correspond the $B=0$ results.} 
\label{fig6}
\end{figure*}

In Fig.~\ref{nl3wr}, the results of the calculations performed for  $B= 4.41 \times 10^{16}$G (left),  $B= 1.3 \times
    10^{17}$G (middle), and  $B= 4.41 \times 10^{17}$G  (right)  without (top)
  and with  (bottom) AMM are shown.  Thin black lines show the
    structure of the $B=0$ calculation. We also include the growth
    rates calculated for the same proton fraction  and magnetic field
    intensity within a dynamical
    calculation \cite{Fang17,Fang17a}. The effect of the magnetic
    field on the structure of the nonhomogeneous matter is different
    from what has been discussed for NL3: in the calculation without AMM, there are no well-separated  nonhomogeneous regions, but there is an extension of the
    nonhomogeneous phase to
    larger densities with respect to the $B=0$ case; in the calculation including AMM, only for the
    strongest field considered a second well disconnected
    region of nonhomogeneous matter does appear. This behavior reflects the pattern of the
    dynamical spinodal unstable regions, which are smaller in
    extension, do not appear for densities very far from the $B=0$
    crust-core transition, and  are characterized by smaller growth rates.
    A second effect is the appearance of new
    geometries:  slabs, tubes and bubbles are also present for the
    highest densities,  as clearly seen in  Fig.~\ref{fig6}.
 In this figure, we identify the geometries of the clusters
      obtained with NL3$\omega\rho$ without (top) and with (bottom)
      AMM.
         For $B=0$, only droplets and
   rods are present, but for  finite $B$,  all geometries occur.
For the two weaker fields, the nonhomogeneous region extends without interruption
   until the crust-core transition. The behavior is more irregular for
   the strongest field,  $B=4.4\times 10^{17}$G: between the
   geometries, some narrow regions of homogeneous matter are present. Also, the extension of the low density region is smaller, and when including AMM,  a
  second region of clusterized matter  emerges at larger densities, with
  slabs, tubes and bubbles.

\section{Conclusions} \label{sec:conclusions}

The present work is an exploratory investigation of the effect of the
magnetic field on the clusterized matter that forms the inner crust of
a neutron star. Previous investigations within a dynamical spinodal
approach have suggested that the magnetic field could give origin to
an extended clusterized region with a crust-core transition occuring
at larger densities \cite{Fang16,Fang17,Fang17a}. Moreover, and depending on the isovector
properties of the EoS, there could exist a detached nonhomogeneous
region above the $B=0$ crust-core transition density.

Our main conclusion is that by taking
 the coexisting phases approach to describe clusterized matter we were able to  confirm the conclusions previously drawn within the
dynamical spinodal calculation, in particular, that the crust-core
transition occurs at larger densities. The effect of the magnetic field is not only reflected on
the extension of the nonhomogeneous phase but also on the appearance
of the different geometries.  Taking two models with a different  density dependence of the
symmetry energy, it was also shown that the effect of the
magnetic field is very sensitive to the behavior of the symmetry energy.

The results of the present study  may help understand the violent
events associated with magnetars. If the inner crust structure is very
sensitive to the magnetic field intensity, stresses will build as the
magnetic field decays due to the change of the equilibrium structure of
the star giving rise to the occurence of fractures \cite{Lander2016,Lander2019,Lander2020}.

Our study requires  further investigation within
self-consistent approaches as, for instance, the Thomas-Fermi approach,  that has been
used to describe clusterized matter of non-magnetized
\cite{Maruyama2005,Avancini-08,Avancini-10,Grill2012,Bao-14} and
magnetized matter \cite{Lima-13,Bao21}, and imposing
  $\beta$-equilibrium. 

In Ref.~\cite{Bao21},  a $\beta$-equilibrium
calculation was carried out. The authors concluded that: larger proton fractions were obtained; no effect on the pasta structures and extension of the
nonhomogeneous matter was obtained for fields $B<10^{17}$
G;  for $B\sim 10^{18}$ G, it was observed that the crust-core
transition occurs at smaller densities and geometries that did not
occur for $B=0$ were present, a  conclusion similar to the one discussed in the
last section for
the strongest field considered.
 In Ref.~\cite{Lima-13}, the
calculation was performed for  proton fractions of 0.1 and
0.3 within the NL3 model. Effects on the radius of the droplets, the
surface tension and the crust-core transition were obtained although a
systematic trend was not seen due to the possible opening of new 
 Landau levels. In both calculations, no nonhomogeneous matter
was reported above the $B=0$ crust-core transition possibly because
this region was not investigated.

 The research was developed within a RMF framework, and restricted to two
models with the same isoscalar properties. We focused on
the effects of a strong magnetic field on the inner crust, and in
particular, the crust-core transition.  Taking the  $B=0$ proton fraction
was an approximation that  allowed us to understand the effect of the $B$ field on the structure of the
inner crust, independently of its effect on the proton fraction. We
think that our qualitative findings should not depend much on the exact value of the proton fraction.

A different framework was applied to study the role of the Landau quantization  of electron motion
on the NS outer crust and the onset of the neutron drip in Refs.~\cite{Chamel2012,Chamel2015} (see also \cite{Blaschke2018}). There, the authors showed, by minimizing the Gibbs
energy and using  atomic masses from the 2012 Atomic Mass Evaluation \cite{AME},
together with masses calculated from the Brussels-Montreal
Hartree-Fock-Bogoliubov nuclear mass model \cite{Goriely2013}, that (1) the
region of the crust  where  neutrons drip
changes with time as the magnetic field decays; (2) the composition of the
outer crust depends on the strength of the magnetic field; (3) less
neutron rich matter is expected for strong magnetic fields; and (4) elastic properties
as the shear modulus are affected by the magnetic field. In these
works, magnetic field intensities up to $\approx 8.8\times 10^{16}$G were
considered. A discussion on the effects of the magnetic
  field on the whole, unified crust, outer and inner, should be carried
  out in the future.

\begin{acknowledgement}

This work was partly supported by the FCT (Portugal) Projects No. UID/FIS/04564/2019,  UID/FIS/04564/2020,  and POCI-01-0145-FEDER-029912, and by PHAROS COST Action CA16214. H.P. acknowledges the grant CEECIND/03092/2017 (FCT, Portugal). 

\end{acknowledgement}

\bibliographystyle{spphys}       % APS-like style for physics

%% Non-BibTeX users please use
%\begin{thebibliography}{}
%%
%% and use \bibitem to create references. Consult the Instructions
%% for authors for reference list style.
%\end{thebibliography}

\end{document}